\documentclass[12pt]{article}\usepackage[]{graphicx}\usepackage[table]{xcolor}
\makeatletter
\def\maxwidth{ %
  \ifdim\Gin@nat@width>\linewidth
    \linewidth
  \else
    \Gin@nat@width
  \fi
}
\makeatother

\definecolor{fgcolor}{rgb}{0.345, 0.345, 0.345}
\newcommand{\hlnum}[1]{\textcolor[rgb]{0.686,0.059,0.569}{#1}}%
\newcommand{\hlsng}[1]{\textcolor[rgb]{0.192,0.494,0.8}{#1}}%
\newcommand{\hlcom}[1]{\textcolor[rgb]{0.678,0.584,0.686}{\textit{#1}}}%
\newcommand{\hlopt}[1]{\textcolor[rgb]{0,0,0}{#1}}%
\newcommand{\hldef}[1]{\textcolor[rgb]{0.345,0.345,0.345}{#1}}%
\newcommand{\hlkwa}[1]{\textcolor[rgb]{0.161,0.373,0.58}{\textbf{#1}}}%
\newcommand{\hlkwb}[1]{\textcolor[rgb]{0.69,0.353,0.396}{#1}}%
\newcommand{\hlkwc}[1]{\textcolor[rgb]{0.333,0.667,0.333}{#1}}%
\newcommand{\hlkwd}[1]{\textcolor[rgb]{0.737,0.353,0.396}{\textbf{#1}}}%

\usepackage{framed}
\makeatletter
\newenvironment{kframe}{%
 \def\at@end@of@kframe{}%
 \ifinner\ifhmode%
  \def\at@end@of@kframe{\end{minipage}}%
  \begin{minipage}{\columnwidth}%
 \fi\fi%
 \def\FrameCommand##1{\hskip\@totalleftmargin \hskip-\fboxsep
 \colorbox{shadecolor}{##1}\hskip-\fboxsep
     \hskip-\linewidth \hskip-\@totalleftmargin \hskip\columnwidth}%
 \MakeFramed {\advance\hsize-\width
   \@totalleftmargin\z@ \linewidth\hsize
   \@setminipage}}%
 {\par\unskip\endMakeFramed%
 \at@end@of@kframe}
\makeatother

\definecolor{shadecolor}{rgb}{.97, .97, .97}
\definecolor{messagecolor}{rgb}{0, 0, 0}
\definecolor{warningcolor}{rgb}{1, 0, 1}
\definecolor{errorcolor}{rgb}{1, 0, 0}
\newenvironment{knitrout}{}{} 

\usepackage{alltt}
\usepackage{amsmath}
\usepackage{graphicx}
\usepackage{url} 
\usepackage{amsfonts}
\usepackage{setspace}
\usepackage{caption}
\usepackage{subcaption}
\usepackage[ruled,noline]{algorithm2e}
\usepackage[table]{xcolor}
\usepackage{mathtools}
\usepackage{authblk}
\usepackage{xspace}
\usepackage{natbib}

\newcommand{\blind}{0}

\newcommand\pkg{\texttt}
\newcommand\code{\texttt}

\newcommand\R{\texttt{R}\xspace}
\newcommand\panelPomp{\texttt{panelPomp}\xspace}
\newcommand\pomp{\texttt{pomp}\xspace}
\newcommand\nimble{\texttt{nimble}\xspace}
\newcommand\mcstate{\texttt{mcstate}\xspace}
\newcommand\spatPomp{\texttt{spatPomp}\xspace}
\newcommand\PanelPOMP{PanelPOMP}
\newcommand\example[1]{ \item #1}

\newcommand\mycolon{{\hspace{0.6mm}:\hspace{0.6mm}}}
\newcommand\np{j}
\newcommand\Np{J}   
\newcommand\nrep{i} 
\newcommand\Nrep{I} 
\newcommand\unit{u} 
\newcommand\Unit{U} 
\renewcommand\time{n}

\newcommand\Nmif{M} 
\newcommand\nmif{m} 
\newcommand\Shared{\Phi}
\newcommand\Specific{\Psi}
\newcommand\shared{\phi}
\newcommand\specific{\psi}
\newcommand\Nshared{A}
\newcommand\nshared{a}
\newcommand\Nspecific{B}
\newcommand\nspecific{b}

\newcommand\argequals{{\,=\,}}
\newcommand\mystretch{\rule[-2mm]{0mm}{5mm} }
\newcommand\myBigStretch{\rule[-3mm]{0mm}{5mm} }
\newcommand\asp{\hspace{6mm}}

\newcommand\RealSpace{\mathbb{R}}
\newcommand\loglik{\lambda}
\newcommand\lik{\ell}
\newcommand\Xspace{{\mathbb X}}
\newcommand\Yspace{{\mathbb Y}}

\newcommand\Thetadim{D}

\newcommand\given{{\, | \,}}
\newcommand\giventh{{\,;\,}}
\newcommand\normal{\mathrm{Normal}}
\newcommand\seq[2]{{#1}\!:\!{#2}}
\newcommand\prob{\mathbb{P}}

\IfFileExists{upquote.sty}{\usepackage{upquote}}{}
\begin{document}

\def\spacingset#1{\renewcommand{\baselinestretch}%
{#1}\small\normalsize} \spacingset{1}



\if0\blind
{
  \title{\bf A tutorial on panel data analysis using partially observed Markov processes via the \R package \panelPomp}

  \author[1]{Carles Bret\'o}
  \author[2]{Jesse Wheeler}
  \author[2]{Aaron A. King}
  \author[2]{Edward L. Ionides}

  \affil[1]{Universitat de Val\`encia}
  \affil[2]{University of Michigan}

  \maketitle
} \fi

\if1\blind
{
  \bigskip
  \bigskip
  \bigskip
  \begin{center}
    {\Large\bf A tutorial on panel data analysis using partially observed Markov processes via the \R package \panelPomp}
\end{center}
  \medskip
} \fi

\bigskip
\begin{abstract}
\noindent The \R package \panelPomp supports analysis of panel data via a general class of partially observed Markov process models (PanelPOMP).
This package tutorial describes how the mathematical concept of a PanelPOMP is represented in the software and demonstrates typical use-cases of \panelPomp.
Monte Carlo methods used for POMP models require adaptation for PanelPOMP models due to the higher dimensionality of panel data.
The package takes advantage of recent advances for PanelPOMP, including an iterated filtering algorithm, Monte Carlo adjusted profile methodology and block optimization methodology to assist with the large parameter spaces that can arise with panel models.
In addition, tools for manipulation of models and data are provided that take advantage of the panel structure.
\end{abstract}

\vfill

\newpage
\spacingset{1.25}

\AtBeginEnvironment{algorithm}{%
  \singlespacing
}

\section{Introduction}
\label{sec:intro}

This tutorial describes a typical use-case of the \panelPomp \R package.
Partially observed Markov process (POMP) models---also known as state-space or hidden Markov models---are useful mathematical tools for modeling non-linear dynamic systems.
POMP models describe a system via an unobserved dynamic model that has the Markov property, coupled with a model for how observations are drawn from the latent process.
Various software packages provide platforms for performing statistical analysis of these systems using POMP models (for instance, \pomp \citep{king16}, \spatPomp \citep{asfaw21}, \nimble \citep{michaud21}, and \mcstate \citep{fitzjohn20}).
However, particular challenges arise when modeling \emph{panel data} via POMP models; these data arise when time series are measured on a collection of independent units.
While each unit may be modeled separately, analyzing the data as a single collection can provide insights into the underlying dynamical system that may not be obtained otherwise.
For instance, each time series in the panel may be too short to infer a complex dynamical model, so that inference on an underlying model must combine information across the units.
\panelPomp is, to our knowledge, the first software package specifically addressing these issues.
The utility of \panelPomp has been demonstrated in several scientific applications \citep{ranjeva17,ranjeva19,wale19,domeyer22,lee20}.

The current version of \panelPomp emphasizes simulation-based methods, also known as plug-and-play methods \citep{breto09,he10}, or likelihood-free methods \citep{marjoram03,sisson07}.
Such methods are applicable to dynamic models for which a simulator is available even when the transition densities are unavailable.
This class of flexible algorithms allow researchers to build their models based on scientific reasoning rather than statistical convenience, typically at the expense of computational efficiency.
In the following sections, this tutorial demonstrates how the \panelPomp package can be used to model nonlinear dynamic systems using plug-and-play methodologies.

\section{PanelPOMP models} \label{sec:PanelPOMP}

Panel data is a collection of multiple time series datasets, each possibly multivariate on its own, where each time series is associated with a unit; the units can represent spatial locations, agents in a system, or other units for which data is collected over time.
For convenience of identifying units in the panel, we use numeric labels $\{1,2,\dots,\Unit\}$, which we also write as $1\mycolon\Unit$.
Time series from each unit may be of different lengths, and so we define $N_\unit$ as the number of measurements collected on unit $\unit$.
The observations are modeled as a realization of a stochastic process $Y_{\unit,1:N_\unit}$, observed at times $t_{\unit,1}<t_{\unit,2}<\dots<t_{\unit,N_\unit}$.
An arbitrary realization of the observable process at time $t_{\unit,n}$ is denoted as $y_{\unit,n}$, and the entire collection of data is written as $y^*_{\unit,1:N_{\unit}} = \{y^*_{\unit,1},\dots,y^*_{\unit,N_\unit}\}$, using the asterisk to differentiate the observed data from an arbitrary realization.
The measurement process is assumed to be dependent on a latent Markov process $\{X_\unit(t),t_{\unit,0}\le t\le t_{\unit,N_\unit}\}$ defined subsequent to an initial time $t_{\unit,0}\le t_{\unit,1}$.
Requiring that $\{X_\unit(t)\}$ and $\{Y_{\unit,i},i\neq n\}$ are independent of $Y_{\unit,n}$ given $X_\unit(t_{\unit,n})$, for each $n\in 1\mycolon N_{\unit}$, completes the partially observed Markov process (POMP) model structure for unit~$\unit$.
For a {\PanelPOMP} we require additionally that all units are modeled as independent.

The latent process can be modeled as either a discrete or continuous time process.
For the continuous time process, the value of the latent states at observation times is of particular interest, so we write $X_{\unit,n}=X_\unit(t_{\unit,n})$.
We suppose that $X_{\unit,n}$ and $Y_{\unit,n}$ take values in arbitrary spaces $\Xspace_{\unit}$ and $\Yspace_{\unit}$ respectively, and that $X_{\unit,0:N_\unit}$ and $Y_{\unit,1:N_\unit}$ have a joint density written as $f_{X_{\unit,0:N_\unit}Y_{\unit,1:N_\unit}}(x_{\unit,0:N_\unit},y_{\unit,1:N_\unit}\giventh\theta)$ with dependence on an unknown real-valued parameter vector $\theta\in\RealSpace^{\Thetadim}$.
The transition density
$f_{X_{\unit,n}|X_{\unit,n-1}}(x_{\unit,n}\given x_{\unit,n-1}\giventh\theta)$
and measurement density
$f_{Y_{\unit,n}|X_{\unit,n}}(y_{\unit,n}\given x_{\unit,n}\giventh\theta)$
are permitted to depend arbitrarily on $u$ and $n$, allowing non-stationary models and the inclusion of covariate time series.
The marginal density of $Y_{\unit,1:N_\unit}$ at $y_{\unit,1:N_\unit}$ is $f_{Y_{\unit,1:N_\unit}}(y_{\unit,1:N_\unit}\giventh\theta)$ and the likelihood function for unit $\unit$ is
$\lik_{\unit}(\theta) = f_{Y_{\unit,1:N_\unit}}(y^*_{\unit,1:N_\unit}\giventh\theta)$.
The likelihood for the entire panel is
$\lik(\theta) = \prod_{\unit=1}^{\Unit} \lik_{\unit}(\theta)$,
and any solution $\hat\theta=\arg\max\lik(\theta)$ is a maximum likelihood estimate (MLE).
The log likelihood is $\loglik(\theta)=\log \lik(\theta)$.

We introduce a structure to the parameter space which is not part of the general definition of a PanelPOMP model, but which is sufficiently common to deserve attention.
Suppose the parameter vector can be written as
$\theta=(\shared,\specific_1,\dots,\specific_\Unit)$,
where
\begin{eqnarray}
\label{eq:b1a}
f_{X_{\unit,\time}\vert X_{\unit,\time-1}}(x_{\unit,\time}\given x_{\unit,\time-1} \giventh \theta)
&=&
f_{X_{\unit,\time}|X_{\unit,\time-1}}(x_{\unit,\time}\given x_{\unit,\time-1} \giventh \shared,\specific_\unit)
\\
\label{eq:b1b}
f_{Y_{\unit,\time}|X_{\unit,\time}}(y_{\unit,\time}\given x_{\unit,\time} \giventh \theta) &=& f_{Y_{\unit,\time}|X_{\unit,\time}}(y_{\unit,\time}\given x_{\unit,\time} \giventh \shared,\specific_\unit)
\\
\label{eq:b1c}
f_{X_{\unit,0}}(x_{\unit,0} \giventh \theta) &=& f_{X_{\unit,0}}(x_{\unit,0} \giventh \shared,\specific_\unit)
\end{eqnarray}
Then, $\specific_{\unit}$ is a vector of \emph{unit-specific} parameters for unit $\unit$, and $\shared$ is a \emph{shared} parameter vector.
We suppose $\shared\in\RealSpace^{\Nshared}$ and $\specific\in\RealSpace^{\Nspecific}$, so the dimension of the parameter vector $\theta$ is $\Thetadim=\Nshared+\Nspecific\Unit$.
The collection of unit-specific parameters can be considered as a $\Nspecific\times\Unit$ matrix $[\specific_{\nspecific,\unit}]$.
Determining which parameters should be modeled as unit-specific and which should be shared is often itself an interesting scientific and statistical question.

\begin{table}[t!]
  \begin{center}
    \begin{tabular}{lll}
      \hline
      Method &Mathematical terminology \\
      \hline
      \code{rprocess} &Simulate from $f_{X_{\unit,n}|X_{\unit,n-1}}( x_{\unit,n} \given x_{\unit,n-1}\giventh \shared,\specific_\unit)$\\
      \code{dprocess} &Evaluate $f_{X_{\unit,n}|X_{\unit,n-1}}( x_{\unit,n} \given x_{\unit,n-1}\giventh \shared,\specific_{\unit})$\\
      \code{rmeasure} &Simulate from $f_{Y_{\unit,n}|X_{\unit,n}}( y_{\unit,n} \given x_{\unit,n}\giventh \shared,\specific_\unit)$\\
      \code{dmeasure} &Evaluate $f_{Y_{\unit,n}|X_{\unit,n}}( y_{\unit,n} \given x_{\unit,n}\giventh \shared,\specific_{\unit})$\\
       \code{rinit} &Simulate from $f_{X_{\unit,0}}( x_{\unit,0} \giventh \shared,\specific_\unit)$\\
    \code{rprior} & Simulate from the prior distribution $\pi(\theta)$ \\
    \code{dprior} & Evaluate the prior density $\pi(\theta)$ \\\hline
    \end{tabular}
  \end{center}
  \caption{Basic model components for \code{pomp} units making up a \code{panelPomp}.
    \label{tab:notation}
  }
\end{table}

The \code{panelPomp} class follows the mathematical structure described above, consisting of a list of \code{pomp} objects together with a specification of shared and unit-specific parameters.
The \code{pomp} objects can be built using the \code{pomp()} constructor function from the \pomp \R package \citep{king16}.
These objects are bound into a \code{panelPomp} object by the constructor function \code{panelPomp()}.
The general framework does not insist that units of a PanelPOMP share observation times or other model features---though any additional shared structure may simplify the specification of the list of constituent \code{pomp} objects.
The specification of \code{pomp} models was discussed by \citet{king16}, and open-source examples are available online for diverse applications including those described in Section~\ref{sec:examples}.
Briefly, the model is specified by writing code to evaluate some or all of the following basic model components described in Table~\ref{tab:notation}.

Algorithms written for \panelPomp may access these functions.
An algorithm is defined to be plug-and-play if it does not require \code{dprocess}.
A \code{pomp} or \code{panelPomp} object does not need to have all the basic computations defined.
In particular, if employing plug-and-play methodology there is no need to specify \code{dprocess}.
Some algorithms may require model components beyond those tabulated above.
For example, the \code{parameter\_trans} function defines parameter transformations which may be carried out to facilitate model fitting by removing boundaries and/or shifting to a natural scale scale for exploring additive perturbations.
In addition, Bayesian methods may call \code{rprior} or \code{dprior} if these have been defined.

A simple example of a {\PanelPOMP} is a stochastic version of the discrete-time Gompertz model for biological population growth \citep{winsor32}.
This model supposes that the density, $X_{\unit,n+1}$, of a population $\unit$ at time $n+1$ depends on the density, $X_{\unit,n}$, at time $n$ according to
\begin{equation}
\label{eq:gompertz1}
X_{\unit,n+1}=\kappa_\unit^{1-e^{-r_\unit}}\,X_{\unit,n}^{e^{-r_\unit}}\,\varepsilon_{\unit,n}.
\end{equation}
In \eqref{eq:gompertz1}, $\kappa_\unit$ is the carrying capacity of population $\unit$, $r_\unit$ is a positive parameter, and $\{\varepsilon_{\unit,n},\unit\in 1\mycolon\Unit,n\in 1\mycolon N_\unit\}$ are independent and identically-distributed lognormal random variables with $\log\varepsilon_{\unit,n}\sim\normal(0,\sigma_{\mathrm{G},\unit}^2)$.
We suppose the population is observed with lognormally distributed errors,
\begin{equation}
\label{eq:gompertz-obs}
\nonumber
\log{Y_{\unit,n}}\;\sim\;\normal\left(\log{X_{\unit,n}},\tau_\unit^2\right).
\end{equation}
This is accessible via
\begin{knitrout}
\definecolor{shadecolor}{rgb}{0.969, 0.969, 0.969}\color{fgcolor}\begin{kframe}
\begin{alltt}
\hldef{gomp} \hlkwb{<-} \hlkwd{panelGompertz}\hldef{(}\hlkwc{N} \hldef{=} \hlnum{100}\hldef{,} \hlkwc{U} \hldef{=} \hlnum{50}\hldef{)}
\end{alltt}
\end{kframe}
\end{knitrout}
\noindent Here, the number of units is \code{length(gomp)=50}.
\panelPomp uses S4 classes \citep{chambers98,genolini08} with \code{gomp} having the base class \code{panelPomp}.
Because data are a key component of PanelPOMP models, the \code{panelGompertz} function first creates a PanelPOMP model, and then generates a dataset by simulating from that model using a reproducible seed specified by the \code{seed} argument to the function.

\begin{figure}
\begin{center}
\resizebox{3in}{!}{\includegraphics{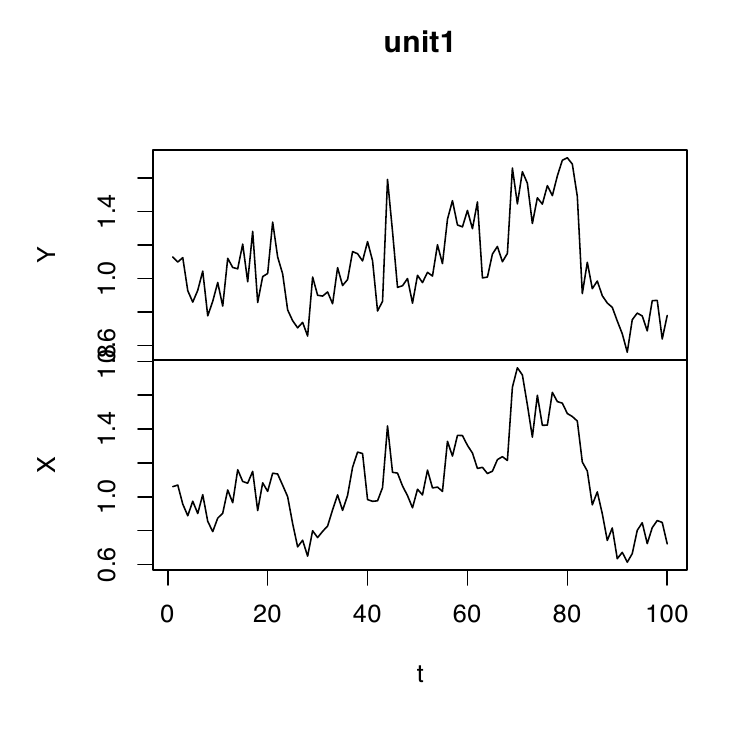}}
\resizebox{3in}{!}{\includegraphics{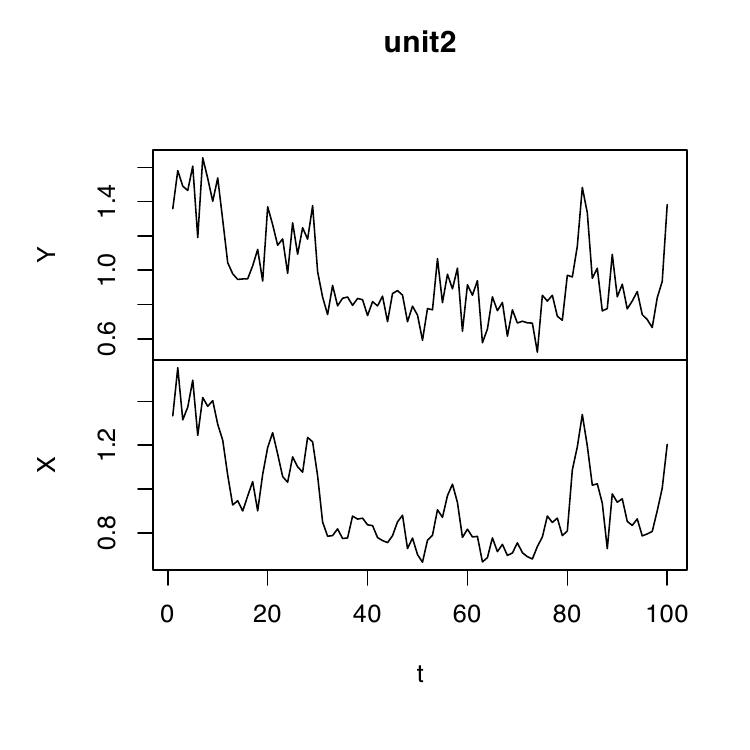}}
\caption{Separate plots produced by \code{plot(gomp[1:2])}.}
\label{fig-plot-gomp}
\end{center}
\end{figure}

Commonly, the first thing to do with a new object is to plot it, and Fig.~\ref{fig-plot-gomp} demonstrates the panelPomp \code{plot} method applied after subsetting \code{gomp}.
The units of a panelPomp do not necessarily share the same variables, so in general a sequence of separate plots is all that can be offered.
It may happen that the units can meaningfully be plotted on the same axis, and that can be achieved by coercing the \code{panelPomp} object to a \code{pompList} and using the \code{plot} method for that class (Fig.~\ref{fig:plot-gomp-pompList}).
A third option is to export the \code{panelPomp} object via \code{as(gomp,"data.frame")} and work this this to produce customized plots.

\begin{knitrout}
\definecolor{shadecolor}{rgb}{0.969, 0.969, 0.969}\color{fgcolor}\begin{figure}

{\centering \includegraphics[width=\maxwidth]{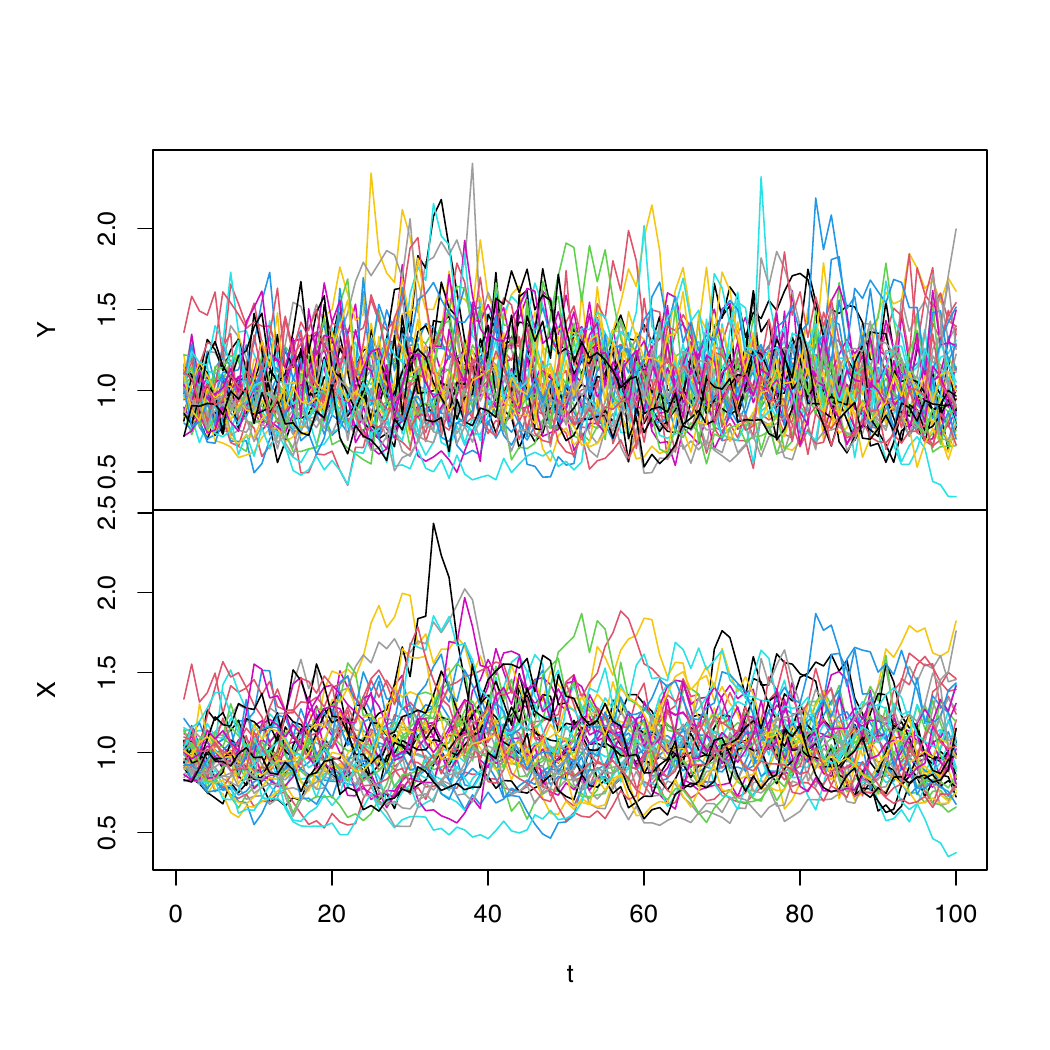} 

}

\caption{Overlayed time series plot using \code{plot(as(gomp,'pompList')).}}\label{fig:plot-gomp-pompList}
\end{figure}

\end{knitrout}

A basic operation is simulation.
We can generate another simulation with the same parameter values:
\begin{knitrout}
\definecolor{shadecolor}{rgb}{0.969, 0.969, 0.969}\color{fgcolor}\begin{kframe}
\begin{alltt}
\hldef{gomp2} \hlkwb{<-} \hlkwd{simulate}\hldef{(gomp)}
\end{alltt}
\end{kframe}
\end{knitrout}
\noindent We can simulate from the same model at new parameter values by giving additional arguments to \code{simulate}.
There are two different representations of parameters within \panelPomp which are convenient in different situations.
The unit-specific parameters are naturally represented as a matrix, set with the \code{specific} argument in the \code{panelPomp} constructor function, with a column for each units and a row for each parameter.
Similarly, the shared parameters are a named vector that can be set using the \code{shared} argument.
Alternatively, we can consider the parameters as a single named vector, with a naming convention that \code{"beta[unit7]"} is the name of the unit-specific parameter \code{beta} for unit $u=7$.
Model parameters can be extracted and set in this vector-format using the functions \code{coef()} and \code{coef()<-}, respectively.
Alternatively, unit-specific and shared parameters can be extracted using functions \code{specific()} and \code{shared()}, and modified using the equivalent setter functions \code{specific()<-} and \code{shared()<-}.
For example, all shared and a subset of unit-specific parameters (from units 1--3) can be extracted in vector format via
\begin{knitrout}
\definecolor{shadecolor}{rgb}{0.969, 0.969, 0.969}\color{fgcolor}\begin{kframe}
\begin{alltt}
\hlkwd{coef}\hldef{(gomp[}\hlnum{1}\hlopt{:}\hlnum{3}\hldef{])}
\end{alltt}
\begin{verbatim}
##          r      sigma   K[unit1] tau[unit1] X.0[unit1]   K[unit2] tau[unit2] 
##        0.1        0.1        1.0        0.1        1.0        1.0        0.1 
## X.0[unit2]   K[unit3] tau[unit3] X.0[unit3] 
##        1.0        1.0        0.1        1.0
\end{verbatim}
\end{kframe}
\end{knitrout}
\noindent or in a list format using the \code{format = `list'} argument
\begin{knitrout}
\definecolor{shadecolor}{rgb}{0.969, 0.969, 0.969}\color{fgcolor}\begin{kframe}
\begin{alltt}
\hlkwd{coef}\hldef{(gomp[}\hlnum{1}\hlopt{:}\hlnum{3}\hldef{],} \hlkwc{format} \hldef{=} \hlsng{'list'}\hldef{)}
\end{alltt}
\begin{verbatim}
## $shared
##     r sigma 
##   0.1   0.1 
## 
## $specific
##     unit1 unit2 unit3
## K     1.0   1.0   1.0
## tau   0.1   0.1   0.1
## X.0   1.0   1.0   1.0
\end{verbatim}
\end{kframe}
\end{knitrout}
\noindent The functions \code{toParamList} and \code{toParamVec} facilitate movement between the vector and list formats
\begin{knitrout}
\definecolor{shadecolor}{rgb}{0.969, 0.969, 0.969}\color{fgcolor}\begin{kframe}
\begin{alltt}
\hlkwd{toParamList}\hldef{(}\hlkwd{coef}\hldef{(gomp[}\hlnum{1}\hlopt{:}\hlnum{3}\hldef{]))}
\end{alltt}
\begin{verbatim}
## $shared
##     r sigma 
##   0.1   0.1 
## 
## $specific
##      unit
## param unit1 unit2 unit3
##   K     1.0   1.0   1.0
##   tau   0.1   0.1   0.1
##   X.0   1.0   1.0   1.0
\end{verbatim}
\end{kframe}
\end{knitrout}

\panelPomp seeks to avoid unnecessary duplication with \pomp.
Thus, \panelPomp requires that \pomp is loaded and builds on existing functionality of \pomp where possible.
In particular, a list of \code{pomp} objects for each unit can be extracted from a \code{panelPomp} object via
\begin{knitrout}
\definecolor{shadecolor}{rgb}{0.969, 0.969, 0.969}\color{fgcolor}\begin{kframe}
\begin{alltt}
\hlkwd{as}\hldef{(panelPompObject,}\hlsng{"list"}\hldef{)}
\end{alltt}
\end{kframe}
\end{knitrout}
Some methods in the \pkg{pomp} package take advantage of a \code{pompList} class which is defined as a list of \code{pomp} objects.
These methods can be accessed via
\begin{knitrout}
\definecolor{shadecolor}{rgb}{0.969, 0.969, 0.969}\color{fgcolor}\begin{kframe}
\begin{alltt}
\hlkwd{as}\hldef{(panelPompObject,}\hlsng{"pompList"}\hldef{)}
\end{alltt}
\end{kframe}
\end{knitrout}
Many critical issues in computational performance of the basic model components for \panelPomp, and utilities for assisting with the user specification of the model, have already received extensive development and testing in the context of \pomp.
All the facilities for constructing POMP models in \pomp are available for constructing the models for each unit in \panelPomp.
This article avoids duplication by referring the reader to \citet{king16} and the \pomp documentation for detailed discussion of constructing \code{pomp} objects.
In Section~\ref{sec:examples}, we identify some case studies that provide useful code for scientific applications of \panelPomp.
In Section~\ref{sec:meth}, we proceed to demonstrate inference for \code{panelPomp} objects, emphasizing methodological issues arising due to the specific requirements of panel data,  in the context of a toy example.

\section{Implementing mechanistic models: case studies}
\label{sec:examples}

Developing mechanistic statistical models for new scientific applications requires identifying the essential variables and their functional relationships, and obtaining a satisfactory description of stochasticity in both the measurement process and the system dynamics.
This challenging but valuable exercise is assisted by simulation-based software that permits implementation of a general class of models.
Once a suitable model is found, it provides a testable benchmark for subsequent investigations of the system under study and other comparable systems.
Published models and methods, equipped with data and reproducible source code, are essential to maintain this progressive development.

Since each unit in a  \code{panelPomp} is itself a \code{pomp}, we refer to \citet{king16} for full details of how these are specified, and we focus on the new issues arising in \code{panelPomp}.
Recall from Section~\ref{sec:PanelPOMP} that a \code{panelPomp} is constructed from the list of constituent \code{pomp} models together with a collection of shared and unit-specific parameters.
A unit-specific parameter named \code{theta} should be called \code{theta} in each constituent \code{pomp}.
Thus, the value \code{theta[unit7]} specific to unit 7 is just passed as \code{theta} when required by the \code{pomp} model representing this unit.

Existing examples of \code{panelPomp} analysis have primarily concerned infectious disease dynamics, a topic that has motivated many advances in inference for partially observed stochastic dynamic systems.
We discuss five of these below.
In addition, the close relationship between \code{panelPomp} and \code{pomp} objects means that \code{panelPomp} model constructions can borrow from the considerable existing resources for \pomp.
The remaining examples give some applications in other domains which have been carried out using \code{pomp} models but have extensions to \code{panelPomp} situations.

\begin{enumerate}
\example{Sexual contacts: behavioral heterogeneity within and between individuals}.
\citet{romero-severson15} developed a PanelPOMP model to investigate a longitudinal prospective survey of sexual contacts, quantifying the roles of behavioral differences between individuals and differences within an individual over time. The \code{contacts()} function in \panelPomp generates one of the models and datasets studied in this paper.
The source code to generate this \code{panelPomp} object is at \url{https://github.com/cbreto/panelPomp}.

\example{Recurrent infection with HPV}.
\citet{ranjeva17} developed a PanelPOMP model to study the strain dynamics of a longitudinal prospective serological survey of human papillomavirus (HPV).
Their \panelPomp code is available at \url{https://github.com/cobeylab/HPV-model}.

\example{Polio: asymptomatic infection and local extinction}.
\citet{martinez-bakker15} developed a POMP model for polio transmission to investigate the pre-vaccination epidemics in USA, fitting all parameters separately for each state.
\citet{breto19} found additional precision in inferences when the states are combined into a PanelPOMP with some shared parameters.
The \code{polio()} function in \panelPomp generates this model, and the source code is at \url{https://github.com/cbreto/panelPomp}.

\example{Age-specific differences in the dynamics of protective immunity to influenza}.
\citet{ranjeva19} developed a PanelPOMP model to interpret longitudinal study of serological measurements on human influenza immunity.
The source code for their \panelPomp model is at \url{https://github.com/cobeylab/Influenza-immune-dynamics}.

\example{The dynamic struggle between malaria and the immune system}.
\citet{wale19} developed a PanelPOMP model to investigate the dynamics of the immune response to malaria, based on flow cytometry time series for a panel of mice under varying treatments. Their \code{panelPomp} source code and data are available at \url{https://doi.org/10.5061/dryad.nk98sf7pk}.

\example{Ecological predator-prey dynamics: consumptive and non-consumptive effects}.
\citet{marino19} developed a stochastic seasonal predator-prey POMP model to investigate the relationship between an abundant zooplankton species, {\it Daphnia mendotae}, and its predator, {\it Bythotrephes longimanus}, in Lake Michigan. The source code for the \code{pomp} analysis is available at \url{https://doi.org/10.5061/dryad.bh688ft}.

\example{Stochastic volatility and financial leverage}.
 \citet{breto14} demonstrated the applicability of plug-and-play methods within \pomp to investigate stochastic volatility in finance using a POMP model for a single index.

\end{enumerate}

Many other POMP models implemented using \code{pomp} are presented at \url{https://kingaa.github.io/pomp/biblio.html}.

\section{Methodology for PanelPOMP models} \label{sec:meth}

All POMP methods can in principle be extended to PanelPOMPs since a PanelPOMP can be written as a POMP.
Three different ways to represent a PanelPOMP as a POMP were identified by \citet{romero-severson15}:
(i) the panels can be concatenated temporally into a long time series; (ii) the panels can be adjoined to form a high-dimensional POMP with a latent state comprised of a vector of latent states for each unit; (iii) time in the POMP representation can correspond to unit, $u$, with a vector valued state representing the full process for this unit.
The existence of these representations does not necessarily imply that POMP methods will be computationally feasible on the resulting PanelPOMP.
In particular, sequential Monte Carlo algorithms can have prohibitive scaling difficulties with the high dimensional latent states that can be involved with representations (ii) and (iii).

Here, we focus on describing and demonstrating the plug-and-play likelihood-based inference workflow used in the scienfic examples of Section~\ref{sec:examples}.
This approach builds on likelihood evaluation via the particle filter using \code{pfilter()} and likelihood maximization via iterated filtering using \code{mif2()}.
These algorithms can be formally justified in terms of representation (i) above \citep{breto19}, though the numerical implementation does not in practice have to explicitly construct the concatenation of the \code{panelPomp} object into a \code{pomp} object.

\subsection{Log likelihood evaluation via panel particle filtering}

The particle filter, also known as sequential Monte Carlo, is a standard tool for log likelihood evaluation on non-Gaussian POMP models.
The log likelihood function is a central component of Bayesian and frequentist inference.
Due to the dynamic independence assumed between units, particle filtering can be carried out separately on each unit.
The \code{pfilter} method for \code{panelPomp} objects is therefore a direct extension of the \code{pfilter} method for \code{pomp} objects from the \pomp package.
Repeating \code{pfilter} is advisable to reduce the Monte Carlo error on the log likelihood evaluation and to quantify this error.
The following code carries out replicated evaluations of the log likelihood of \code{gomp}, taking advantage of multicore computation.
The Gompertz model is a convenient for testing methodology for nonlinear non-Gaussian models since it has a logarithmic transformation to a linear Gaussian process and therefore the exact likelihood is computable by the Kalman filter \citep{king16}.

\begin{knitrout}
\definecolor{shadecolor}{rgb}{0.969, 0.969, 0.969}\color{fgcolor}\begin{kframe}
\begin{alltt}
\hldef{pf_results} \hlkwb{<-} \hlkwd{foreach}\hldef{(}\hlkwc{i}\hldef{=}\hlnum{1}\hlopt{:}\hlnum{10}\hldef{)} \hlopt{%dopar%} \hlkwd{pfilter}\hldef{(gomp,}
  \hlkwc{Np}\hldef{=}\hlkwd{switch}\hldef{(run_level,}\hlnum{10}\hldef{,}\hlnum{200}\hldef{,}\hlnum{1000}\hldef{))}
\end{alltt}
\end{kframe}
\end{knitrout}

This took 0.05 minutes using 4 cores, resulting in a list of objects of class \code{pfilterd.ppomp}.
We can use \code{logLik} to extract the Monte Carlo likelihood esimate $\loglik^{[\nrep]}$ for each replicate $\nrep$, and \code{unitlogLik}  to extract the vector of component Monte Carlo likelihood esimates $\loglik^{[\nrep]}_{\unit}$ for each unit $\unit=1,\dots,\Unit$, where $\loglik^{[\nrep]}=\sum_{\unit=1}^{\Unit}\loglik^{[\nrep]}_{\unit}$.
For a POMP model, replicated particle filter likelihood evaluations are usually averaged on the natural scale, rather than the log scale, to take advantage of the unbiasedness of the particle filter likelihood estimate.
Thus, we have
\[
\hat\loglik_1 = \log \frac{1}{\Nrep}\sum_{\nrep=1}^{\Nrep}
\exp
\left\{
  \sum_{\unit=1}^{\Unit} \loglik^{[\nrep]}_{\unit}
\right\}
\]
which can be implemented as
\begin{knitrout}
\definecolor{shadecolor}{rgb}{0.969, 0.969, 0.969}\color{fgcolor}\begin{kframe}
\begin{alltt}
\hldef{lambda_1} \hlkwb{<-} \hlkwd{logmeanexp}\hldef{(}\hlkwd{sapply}\hldef{(pf_results,logLik),}\hlkwc{se}\hldef{=}\hlnum{TRUE}\hldef{)}
\end{alltt}
\end{kframe}
\end{knitrout}
giving $\hat\loglik_1=2066.5$ with a jack-knife standard error of $0.7$.
Taking advantage of the independence of the units in the panel structure, \citet{breto19} showed it is preferable to average the replicates of marginal likelihood for each unit before taking a product over units.
This corresponds to
\[
\hat\loglik_2 = \log \prod_{\unit=1}^\Unit \frac{1}{\Nrep}\sum_{\nrep=1}^{\Nrep}  \exp \big\{ \hat \loglik^{[\nrep]}_{\unit} \big\}
\]
which can be implemented as
\begin{knitrout}
\definecolor{shadecolor}{rgb}{0.969, 0.969, 0.969}\color{fgcolor}\begin{kframe}
\begin{alltt}
\hldef{lambda_2} \hlkwb{<-} \hlkwd{panel_logmeanexp}\hldef{(}\hlkwd{sapply}\hldef{(pf_results,unitlogLik),}
  \hlkwc{MARGIN}\hldef{=}\hlnum{1}\hldef{,}\hlkwc{se}\hldef{=}\hlnum{TRUE}\hldef{)}
\end{alltt}
\end{kframe}
\end{knitrout}
giving $\hat\loglik_2=2067.9$ with a jack-knife standard error of $1.1$.
For this model, a Kalman filter likelihood evaluation gives an exact answer, $\loglik=2068.2$.

\subsection{Maximum likelihood estimation via Panel Iterated Filtering} \label{sec:pif}

Iterated filtering algorithms carry out repeated particle filtering operations on an extended version of the model that includes time-varying perturbations of parameters.
At each iteration, the magnitude of the perturbations is decreased, and in a suitable limit the algorithm approaches a local maximum of the likelihood function.
The IF2 iterated filtering algorithm \citep{ionides15} has been used for likelihood-based inference on various POMP models arising in epidemiology and ecology \citep[reviewed by][]{breto18}, superseding the previous IF1 algorithm of \citet{ionides06}.
IF2 is implemented in \pomp as the \code{mif2} method for class \code{pomp}.
A panel iterated filtering (PIF) algorithm, extending IF2 to panel data, was developed by \citet{breto19}.
An implementation of PIF in \panelPomp is provided by the \code{mif2} method for class \code{panelPomp}, following the pseudocode in Algorithm~\ref{alg:pif}.
The pseudocode in  Algorithm~\ref{alg:pif} sometimes omits explicit specification of ranges over which variables are to be computed when this is apparent from the context:
it is understood that $\np$ takes values in $\seq{1}{\Np}$, $\nshared$ in
$\seq{1}{\Nshared}$ and $\nspecific$ in $\seq{1}{\Nspecific}$.
The $N[0,1]$ notation corresponds to the construction of independent standard normal random variables, leading to to Gaussian perturbations of parameters on a transformed scale.
These perturbations could follow an arbitrary distribution within the theoretical frameworks of IF2 and PIF.

\begin{algorithm}[t!]
  \caption{
    \texttt{mif2$\big($pp, Nmif{\argequals}$M$, Np{\argequals}$J$,
start{\argequals}$(\shared^0_{\nshared},\specific^0_{\nspecific,\unit})$,
rw\_sd{\argequals}$(\sigma^\Shared_{\nshared,\time},\sigma^\Specific_{\nspecific,\unit,\time})$,
cooling.factor.50{\argequals}$\rho^{50}\big)$}, where \code{pp} is a \code{panelPomp} object containing data and defined \code{rprocess}, \code{dmeasure}, \code{rinit} and \code{partrans} components.
    \label{alg:pif}
    }
\noindent\begin{tabular}{ll}
{\bf input:}\rule[-1.5mm]{0mm}{6mm}
& Data, $y_{\unit,n}^*$, $u$ in $\seq{1}{U}$, $n$ in $\seq{1}{N}$\\
& Simulator of initial density, $f_{X_{\unit,0}}(x_{\unit,0}\giventh \shared,\specific_{\unit})$ \\
& Simulator of transition density, $f_{X_{\unit,n}|X_{\unit,n-1}}(x_{\unit,n}\given x_{\unit,n-1}\giventh \shared,\specific_{\unit})$ \\
& Evaluator of measurement density, $f_{Y_{\unit,n}|X_{\unit,n}}(y_{\unit,n}\given x_{\unit,n}\giventh\shared,\specific_{\unit})$ \\
& Number of particles, $J$, and number of iterations, $\Nmif$\\
& Starting shared parameter swarm, $\Shared^0_{\nshared,\np}=\shared^0_{\nshared}$, $a$ in $\seq{1}{A}$, $j$ in $\seq{1}{J}$\\
& Starting unit-specific parameter swarm, $\Specific^0_{\nspecific,\unit,\np}=\specific^0_{\nspecific,\unit}$,  $b$ in  $\seq{1}{B}$, $j$ in $\seq{1}{J}$\\
& Random walk intensities,
$\sigma^\Shared_{\nshared,\time}$ and $\sigma^\Specific_{\nspecific,\unit,\time}$ \\
& Parameter transformations, $h^{\Shared}_{\nshared}$ and $h^{\Specific}_{\nspecific}$, with inverses
 $\big(h^{\Shared}_{\nshared}\big)^{-1}$ and $\big(h^{\Specific}_{\nspecific}\big)^{-1}$
\\
{\bf output:}\rule[-1.5mm]{0mm}{6mm}
& Final parameter swarm, $\Shared^M_{\nshared,\np}\;$ and $\;\Specific^M_{\nspecific,\unit,\np}$
\rule[-2mm]{0mm}{5mm}
\end{tabular}

\hrule

\noindent\begin{tabular}{l}
For $\nmif$ in $1\mycolon M$\rule[0mm]{0mm}{5mm}\\
\asp $\Shared^m_{\nshared,0,\np}=\Shared^{m-1}_{\nshared,\np}$
\mystretch\\
\asp For $\unit$ in $1\mycolon \Unit$\\
\asp \asp $\Shared^{F,m}_{\nshared,\unit,0,j} =\big(h^{\Shared}_{\nshared}\big)^{-1}
  \left(
  h^\Shared_\nshared \big(\Shared^{m}_{\nshared,\unit-1,\np}\big)+
    \rho^{m}\sigma^{\Shared}_{\nshared,0} Z^{\Shared,m}_{\nshared,\unit,0,\np}
  \right)$ for $ Z^{\Shared,m}_{\nshared,\unit,0,\np}\sim N[0,1]$ \myBigStretch\\
\asp \asp $\Specific^{F,m}_{\nspecific,\unit,0,j} =\big(h^{\Specific}_{\nspecific}\big)^{-1}
  \left(
  h^\Specific_\nspecific \big(\Specific^{m-1}_{\nspecific,\unit,\np}\big)+
    \rho^{m}\sigma^{\Specific}_{\nspecific,u,0} Z^{\Specific,m}_{\nspecific,\unit,0,\np}
  \right)$ for $ Z^{\Specific,m}_{\nspecific,\unit,0,\np}\sim N[0,1]$ \myBigStretch\\
\asp \asp $X_{\unit,0,j}^{F,m}\sim f_{X_{\unit,0}}
  \left(
    x_{\unit,0} \; \giventh \; \Shared^{F,m}_{\nshared,\unit,0,j} ,
    \Specific^{F,m}_{\nspecific,\unit,0,j}
  \right)$ \myBigStretch\\
\asp \asp For $n$ in $1\mycolon N_{\unit}$\\
\asp \asp \asp $\Shared^{P,m}_{\nshared,\unit,\time,j} =\big(h^{\Shared}_{\nshared}\big)^{-1}
  \left(
  h^\Shared_\nshared \big(\Shared^{F,m}_{\nshared,\unit,\time-1,\np}\big)+
    \rho^{m}\sigma^{\Shared}_{\nshared,\time} Z^{\Shared,m}_{\nshared,\unit,\time,\np}
  \right)$ for $ Z^{\Shared,m}_{\nshared,\unit,\time,\np}\sim N[0,1]$ \myBigStretch\\
\asp \asp \asp $\Specific^{P,m}_{\nspecific,\unit,\time,j} =\big(h^{\Specific}_{\nspecific}\big)^{-1}
  \left(
  h^\Specific_\nspecific \big(\Specific^{F,m}_{\nspecific,\unit,\time-1,\np}\big)+
    \rho^{m}\sigma^{\Specific}_{\nspecific,u,\time} Z^{\Specific,m}_{\nspecific,\unit,\time,\np}
  \right)$ for $ Z^{\Specific,m}_{\nspecific,\unit,\time,\np}\sim N[0,1]$ \myBigStretch\\
\asp \asp \asp $X_{\unit,n,j}^{P,m}\sim f_{X_{\unit,n}|X_{\unit,n-1}}
   \left(
     x_{\unit,n} \; \big| \; X^{F,m}_{\unit,n-1,j} \; \giventh \;
     \Shared^{P,m}_{\nshared,\unit,\time,j} \, ,
     \Specific^{P,m}_{\nspecific,\unit,\time,j}
   \right)$  \myBigStretch\\
\asp \asp \asp $w_{\unit,n,j}^m = f_{Y_{\unit,n}|X_{\unit,n}}
    \left(y^*_{\unit,n} \; \big| \; X_{\unit,n,j}^{P,m} \; \giventh \;
       \Shared^{P,m}_{\nshared,\unit,\time,j} \, ,
       \Specific^{P,m}_{\nspecific,\unit,\time,j}
    \right)$  \myBigStretch\\
\asp \asp \asp Draw $k_{1:J}$ with $\prob(k_j=i)=  w_{\unit,n,i}^m\Big/\sum_{q=1}^J w_{\unit,n,q}^m$ \myBigStretch \\
\asp \asp \asp $\Shared^{F,m}_{\nshared,\unit,\time,\np}=\Shared^{P,m}_{\nshared,\unit,\time,k_{\np}}$,  $\;\; \Specific^{F,m}_{\nspecific,\unit,\time,\np}=\Specific^{P,m}_{\nspecific,\unit,\time,k_{\np}} \;$
and $\; X^{F,m}_{\unit,n,j}=X^{P,m}_{\unit,n,k_j}$   \mystretch\\
\asp \asp End For \\ 
\asp \asp  $\Shared^{m}_{\nshared,\unit,\np}=\Shared^{F,m}_{\nshared,\unit,N_{\unit},\np}$ and
           $\Specific^{m}_{\nspecific,\unit,\np}=\Specific^{F,m}_{\nspecific,\unit,N_{\unit},\np}$
 \\
\asp End For \\ 
\asp $\Shared^{m}_{\nshared,\np}=\Shared^m_{\nshared,\Unit,\np}$
\\
End For \\ 
\end{tabular}
\end{algorithm}

At a conceptual level, the PIF algorithm has an evolutionary analogy: successive iterations mutate parameters and select among the fittest outcomes measured by Monte Carlo likelihood evaluation.
The theory allows considerable flexibility in how the parameters are perturbed, but Gaussian perturbations on an appropriate scale are typically adequate.
Most often, the perturbation parameters
$\sigma^\Shared_{\nshared,\time}$ and $\sigma^\Specific_{\nspecific,\unit,\time}$ in Algorithm~\ref{alg:pif} will not depend on $\time$.
For parameters set to have uncertainty on a unit scale, the value 0.02 demonstrated here has been commonly used.
The help documentation on the \code{rw\_sd} argument gives instruction on using additional structure should it become necessary.

For positive parameters, a logarithmic transform can achieve both tasks of removing the boundary and placing uncertainty on a unit scale.
For the panel Gompertz model, all the parameters are non-negative valued, and so the \code{panelGompertz()} code calls \code{panelPomp} with an argument
\begin{knitrout}
\definecolor{shadecolor}{rgb}{0.969, 0.969, 0.969}\color{fgcolor}\begin{kframe}
\begin{alltt}
\hldef{partrans}\hlkwb{=}\hlkwd{parameter_trans}\hldef{(}\hlkwc{log}\hldef{=}\hlkwd{c}\hldef{(}\hlsng{"K"}\hldef{,}\hlsng{"r"}\hldef{,}\hlsng{"sigma"}\hldef{,}\hlsng{"tau"}\hldef{,}\hlsng{"X.0"}\hldef{))}
\end{alltt}
\end{kframe}
\end{knitrout}
Inference methodology can call \code{partrans(...,dir="toEst")} to work with parameters on a suitable scale, usually one where additive variation is meaningful.
The methodology can revert to the original parameterization, presumably chosen to be scientifically convenient or meaningful, using  \code{partrans(...,dir="fromEst")}.
Thus, a user who does not have to look `under the hood' never has to be directly concerned with parameters on the transformed scale, beyond assigning the transformation.

For Monte Carlo maximization, replication from diverse starting points is recommended.
We demonstrate such a maximization search on \code{gomp}.
For simplicity, we fix $K_{\unit}=1$ and the initial condition $X_{\unit,0}=1$, maximizing over two shared parameters, $r$ and $\sigma$, and one unit-specific parameter $\tau_\unit$.
To define the diverse starting points, we make uniform draws from a specified box.
We are not promising that the search will stay within this box, and indeed we should be alert to the possibility that the data lead us elsewhere.
However, if replicated searches started from this box reliably reach a consensus, we claim we have carefully investigated this part of parameter space.
A larger box leads to greater confidence that the relevant part of the parameter space has been searched, at the expense of requiring additional work.
The \code{runif\_panel\_design} function constructs a matrix of random draws from the box.
\begin{knitrout}
\definecolor{shadecolor}{rgb}{0.969, 0.969, 0.969}\color{fgcolor}\begin{kframe}
\begin{alltt}
\hldef{starts} \hlkwb{<-} \hlkwd{runif_panel_design}\hldef{(}
  \hlkwc{lower} \hldef{=} \hlkwd{c}\hldef{(}\hlsng{'r'} \hldef{=} \hlnum{0.05}\hldef{,} \hlsng{'sigma'} \hldef{=} \hlnum{0.05}\hldef{,} \hlsng{'tau'} \hldef{=} \hlnum{0.05}\hldef{,} \hlsng{'K'} \hldef{=} \hlnum{1}\hldef{,} \hlsng{'X.0'} \hldef{=} \hlnum{1}\hldef{),}
  \hlkwc{upper} \hldef{=} \hlkwd{c}\hldef{(}\hlsng{'r'} \hldef{=}  \hlnum{0.2}\hldef{,} \hlsng{'sigma'} \hldef{=}  \hlnum{0.2}\hldef{,} \hlsng{'tau'} \hldef{=}  \hlnum{0.2}\hldef{,} \hlsng{'K'} \hldef{=} \hlnum{1}\hldef{,} \hlsng{'X.0'} \hldef{=} \hlnum{1}\hldef{),}
  \hlkwc{specific_names} \hldef{=} \hlkwd{c}\hldef{(}\hlsng{'K'}\hldef{,} \hlsng{'tau'}\hldef{,} \hlsng{'X.0'}\hldef{),}
  \hlkwc{unit_names} \hldef{=} \hlkwd{names}\hldef{(gomp),}
  \hlkwc{nseq}\hldef{=}\hlkwd{switch}\hldef{(run_level,}\hlnum{2}\hldef{,}\hlnum{4}\hldef{,}\hlnum{6}\hldef{)}
\hldef{)}
\end{alltt}
\end{kframe}
\end{knitrout}

We then carry out a search from each starting point:

\begin{knitrout}
\definecolor{shadecolor}{rgb}{0.969, 0.969, 0.969}\color{fgcolor}\begin{kframe}
\begin{alltt}
\hldef{mif_results} \hlkwb{<-} \hlkwd{foreach}\hldef{(}\hlkwc{start}\hldef{=}\hlkwd{iter}\hldef{(starts,}\hlsng{"row"}\hldef{))} \hlopt{%dopar%} \hldef{\{}
  \hlkwd{mif2}\hldef{(gomp,} \hlkwc{start}\hldef{=}\hlkwd{unlist}\hldef{(start),}
    \hlkwc{Nmif} \hldef{=} \hlkwd{switch}\hldef{(run_level,}\hlnum{2}\hldef{,}\hlnum{20}\hldef{,}\hlnum{150}\hldef{),}
    \hlkwc{Np} \hldef{=} \hlkwd{switch}\hldef{(run_level,}\hlnum{10}\hldef{,}\hlnum{500}\hldef{,}\hlnum{1500}\hldef{),}
    \hlkwc{cooling.fraction.50}\hldef{=}\hlnum{0.5}\hldef{,}
    \hlkwc{cooling.type}\hldef{=}\hlsng{"geometric"}\hldef{,}
    \hlkwc{transform}\hldef{=}\hlnum{TRUE}\hldef{,}
    \hlkwc{rw.sd}\hldef{=}\hlkwd{rw_sd}\hldef{(}\hlkwc{r}\hldef{=}\hlnum{0.02}\hldef{,}\hlkwc{sigma}\hldef{=}\hlnum{0.02}\hldef{,}\hlkwc{tau}\hldef{=}\hlnum{0.02}\hldef{)}
  \hldef{)}
\hldef{\}}
\end{alltt}
\end{kframe}
\end{knitrout}

This took 15.0 minutes using 4 cores, producing a list of objects of class \code{mifd.ppomp}.
The algorithmic parameters are very similar to those of the \code{mif2} method for class \code{pomp}.
The perturbations, determined by the \code{rw\_sd} argument, may be a list giving separate instructions for each unit.
When only one specification for a unit-specific parameter is given (as we do for $K_\unit$ here) the same perturbation is used for all units.

We can check on convergence of the searches, and possibly diagnose improvements in the choices of algorithmic parameters, by consulting trace plots of the searches available via the \code{traces} method for class \code{mifd.ppomp}.
This follows recommendations by \citep{ionides06} and \citep{king16}.

An issue characteristic of PanelPOMP models is using the panel structure to facilitate the large number of parameters arising when unit-specific parameters are specified for a large number of units.
For a fixed value of the shared parameters, the likelihood of the unit-specific parameters factorizes over the units.
The factorized likelihood can be maximized separately over each unit, replacing a challenging high-dimensional problem with many relatively routine low-dimensional problems.
This suggests a block maximization strategy where unit-specific parameters for each unit are maximized as a block.
\citet{breto19} used a simple block strategy where a global search over all parameters is followed by a block maximization over units for unit-specific parameters.
We demonstrate this here, refining each of the maximization replicates above.
The following function carries out a maximization search of unit-specific parameters for a single unit.
The call to \code{mif2} takes advantage of argument recycling: all algorithmic parameters are re-used from the construction of \code{mifd\_gomp} except for the respecified random walk standard deviations which ensures that only the unit-specific parameters are perturbed.

\begin{knitrout}
\definecolor{shadecolor}{rgb}{0.969, 0.969, 0.969}\color{fgcolor}\begin{kframe}
\begin{alltt}
\hldef{mif_unit} \hlkwb{<-} \hlkwa{function}\hldef{(}\hlkwc{unit}\hldef{,}\hlkwc{mifd_gomp}\hldef{,}\hlkwc{reps}\hldef{=}\hlkwd{switch}\hldef{(run_level,}\hlnum{2}\hldef{,}\hlnum{4}\hldef{,}\hlnum{6}\hldef{))\{}
  \hldef{unit_gomp} \hlkwb{<-} \hlkwd{unit_objects}\hldef{(mifd_gomp)[[unit]]}
  \hldef{mifs} \hlkwb{<-} \hlkwd{replicate}\hldef{(}\hlkwc{n}\hldef{=reps,}\hlkwd{mif2}\hldef{(unit_gomp,}\hlkwc{rw.sd}\hldef{=}\hlkwd{rw_sd}\hldef{(}\hlkwc{tau}\hldef{=}\hlnum{0.02}\hldef{)))}
  \hldef{best} \hlkwb{<-} \hlkwd{which.max}\hldef{(}\hlkwd{sapply}\hldef{(mifs,logLik))}
  \hlkwd{coef}\hldef{(mifs[[best]])[}\hlsng{"tau"}\hldef{]}
\hldef{\}}
\end{alltt}
\end{kframe}
\end{knitrout}

Now we apply this block maximization to find updated unit-specific parameters for each replicate, and we insert these back into the \code{panelPomp}

\begin{knitrout}
\definecolor{shadecolor}{rgb}{0.969, 0.969, 0.969}\color{fgcolor}\begin{kframe}
\begin{alltt}
\hldef{mif_block} \hlkwb{<-} \hlkwd{foreach}\hldef{(}\hlkwc{mf}\hldef{=mif_results)} \hlopt{%dopar%} \hldef{\{}
  \hldef{mf}\hlopt{@}\hlkwc{specific}\hldef{[}\hlsng{"tau"}\hldef{,]} \hlkwb{<-} \hlkwd{sapply}\hldef{(}\hlnum{1}\hlopt{:}\hlkwd{length}\hldef{(mf),mif_unit,}\hlkwc{mifd_gomp}\hldef{=mf)}
  \hldef{mf}
\hldef{\}}
\end{alltt}
\end{kframe}
\end{knitrout}

This took 0.8 minutes.

We expect Monte Carlo estimates of the maximized log likelihood functions to fall below the actual (usually unknown) value.
This is in part because imperfect maximization can only reduce the maximized likelihood, and in part a consequence of Jensen's inequality applied to the likelihood evaluation: the unbiased SMC likelihood evaluation has a negative bias on estimation of the log likelihood.

\subsection{Monte Carlo profile likelihood}

The profile likelihood function is constructed by fixing one focal parameter at a range of values and then maximizing the likelihood over all other parameters for each value of the focal parameter.
Constructing a profile likelihood function has several practical advantages.
\begin{enumerate}
\item Evaluations at neighboring values of the focal parameter provide additional Monte Carlo replication. Typically, the true profile log likelihood is smooth, and asymptotically close to quadratic under regularity conditions, so deviations from a smooth fitted line can be interpreted as Monte Carlo error.
\item Large-scale features of the profile likelihood reveal a region of the parameter space outside which the model provides a poor explanation of the data.
\item Co-plots, showing how the values of other maximized parameters vary along the profile, may provide insights into parameter tradeoffs implied by the data.
\item The smoothed Monte Carlo profile log likelihood can be used to construct an approximate 95\% confidence interval.
The resulting confidence interval can be properly adjusted to accommodate both statistical and Monte Carlo uncertainty \citep{ionides17}.
\end{enumerate}

Once we have code for maximizing the likelihood, only minor adaptation is needed to carry out the maximizations for a profile.
The \code{runif\_panel\_design} generating the starting values is replaced by a call to \code{profile\_design}, which assigns the focal parameter to a grid of values and randomizes the remaining parameters.
The random walk standard deviation for the focal parameter is unassigned, which leads it to be set to zero and therefore the parameter remains fixed during the maximization process.
The following code combines the joint and block maximizations developed above.

\begin{knitrout}
\definecolor{shadecolor}{rgb}{0.969, 0.969, 0.969}\color{fgcolor}\begin{kframe}
\begin{alltt}
\hlcom{# Names of the estimated parameters}
\hldef{estimated} \hlkwb{<-} \hlkwd{c}\hldef{(}
  \hlsng{"r"}\hldef{,} \hlsng{"sigma"}\hldef{,} \hlkwd{paste0}\hldef{(}\hlsng{"tau[unit"}\hldef{,} \hlnum{1}\hlopt{:}\hlkwd{length}\hldef{(gomp),} \hlsng{"]"}\hldef{)}
\hldef{)}

\hlcom{# Names of the fixed parameters (not estimated)}
\hldef{fixed} \hlkwb{<-} \hlkwd{names}\hldef{(}\hlkwd{coef}\hldef{(gomp))[}\hlopt{!}\hlkwd{names}\hldef{(}\hlkwd{coef}\hldef{(gomp))} \hlopt{%in%} \hldef{estimated]}

\hldef{profile_starts} \hlkwb{<-} \hlkwd{profile_design}\hldef{(}
  \hlkwc{r} \hldef{=} \hlkwd{seq}\hldef{(}\hlnum{0.05}\hldef{,} \hlnum{0.2}\hldef{,} \hlkwc{length} \hldef{=} \hlkwd{switch}\hldef{(run_level,} \hlnum{10}\hldef{,} \hlnum{10}\hldef{,} \hlnum{20}\hldef{)),}
  \hlkwc{lower} \hldef{=} \hlkwd{c}\hldef{(}\hlkwd{coef}\hldef{(gomp)[estimated]} \hlopt{/} \hlnum{2}\hldef{,} \hlkwd{coef}\hldef{(gomp)[fixed])[}\hlopt{-}\hlnum{1}\hldef{],}
  \hlkwc{upper} \hldef{=} \hlkwd{c}\hldef{(}\hlkwd{coef}\hldef{(gomp)[estimated]} \hlopt{*} \hlnum{2}\hldef{,} \hlkwd{coef}\hldef{(gomp)[fixed])[}\hlopt{-}\hlnum{1}\hldef{],}
  \hlkwc{nprof} \hldef{=} \hlnum{2}\hldef{,} \hlkwc{type} \hldef{=} \hlsng{"runif"}
\hldef{)}

\hldef{profile_results} \hlkwb{<-} \hlkwd{foreach}\hldef{(}\hlkwc{start}\hldef{=}\hlkwd{iter}\hldef{(profile_starts,}\hlsng{"row"}\hldef{))} \hlopt{%dopar%} \hldef{\{}
  \hldef{mf} \hlkwb{<-} \hlkwd{mif2}\hldef{(}
    \hldef{mif_results[[}\hlnum{1}\hldef{]],}
    \hlkwc{start} \hldef{=} \hlkwd{unlist}\hldef{(start),}
    \hlkwc{rw.sd} \hldef{=} \hlkwd{rw_sd}\hldef{(}\hlkwc{sigma} \hldef{=} \hlnum{0.02}\hldef{,} \hlkwc{tau} \hldef{=} \hlnum{0.02}\hldef{))}
  \hldef{mf}\hlopt{@}\hlkwc{specific}\hldef{[}\hlsng{"tau"}\hldef{, ]} \hlkwb{<-} \hlkwd{sapply}\hldef{(}\hlnum{1}\hlopt{:}\hlkwd{length}\hldef{(mf), mif_unit,}\hlkwc{mifd_gomp} \hldef{= mf)}
  \hldef{mf}
\hldef{\}}
\end{alltt}
\end{kframe}
\end{knitrout}

The profile searches took 68.2 minutes.
However, we are not quite done gathering the results for the profile.
The perturbed filtering carried out by \code{mif2} leads to an approximate likelihood evaluation, but for our main results it is better to re-evaluate the likelihood without perturbations.
Also, replication is recommended to reduce and quantify Monte Carlo error.
We do this, and tabulate the results.

\begin{knitrout}
\definecolor{shadecolor}{rgb}{0.969, 0.969, 0.969}\color{fgcolor}\begin{kframe}
\begin{alltt}
\hldef{profile_table} \hlkwb{<-} \hlkwd{foreach}\hldef{(}\hlkwc{mf}\hldef{=profile_results,}\hlkwc{.combine}\hldef{=rbind)} \hlopt{%dopar%} \hldef{\{}
  \hldef{LL} \hlkwb{<-} \hlkwd{replicate}\hldef{(}\hlkwd{switch}\hldef{(run_level,}\hlnum{2}\hldef{,}\hlnum{5}\hldef{,}\hlnum{10}\hldef{),}
    \hlkwd{logLik}\hldef{(}\hlkwd{pfilter}\hldef{(mf,}\hlkwc{Np}\hldef{=}\hlkwd{switch}\hldef{(run_level,}\hlnum{10}\hldef{,}\hlnum{500}\hldef{,}\hlnum{2500}\hldef{)))}
  \hldef{)}
  \hldef{LL} \hlkwb{<-} \hlkwd{logmeanexp}\hldef{(LL,}\hlkwc{se}\hldef{=}\hlnum{TRUE}\hldef{)}
  \hlkwd{data.frame}\hldef{(}\hlkwd{t}\hldef{(}\hlkwd{coef}\hldef{(mf)),}\hlkwc{loglik}\hldef{=LL[}\hlnum{1}\hldef{],}\hlkwc{loglik.se}\hldef{=LL[}\hlnum{2}\hldef{])}
\hldef{\}}
\end{alltt}
\end{kframe}
\end{knitrout}

The likelihood evaluations took 3.2 minutes.
It is appropriate to spend comparable time evaluating the likelihood to the time spent maximizing it: a high quality maximization without high quality likelihood evaluation is hard to interpret, whereas good evaluations of the likelihood in a vicinity of the maximum can inform about the shape of the likelihood surface in this region which may be as relevant as knowing the exact maximum.

The Monte Carlo adjusted profile (MCAP) approach of \citet{ionides17} is implemented by the \code{mcap()} function in \pomp.
This function constructs a smoothed profile likelihood, by application of the \code{loess} smoother.
It computes a local quadratic approximation that is used to derive an extension to the classical profile likelihood confidence interval that makes allowance for Monte Carlo error in the calculation of the profile points.
Theoretically, an MCAP procedure can obtain statistically efficient confidence intervals even when the Monte Carlo error in the profile likelihood is asymptotically growing and unbounded \citep{ning21}.
Log likelihood evaluation has negative bias, as a consequence of Jensen's inequality for an unbiased likelihood estimate.
This bias produces a vertical shift in the estimated profile, which fortunately does not have consequence for the confidence interval if the bias is slowly varying.

The profile points evaluated above, and stored in \code{profile\_table}, can be used to compute a 95\% MCAP confidence interval as follows:
\begin{knitrout}
\definecolor{shadecolor}{rgb}{0.969, 0.969, 0.969}\color{fgcolor}\begin{kframe}
\begin{alltt}
\hldef{gomp_mcap} \hlkwb{<-} \hldef{pomp}\hlopt{::}\hlkwd{mcap}\hldef{(}\hlkwc{logLik}\hldef{=profile_table}\hlopt{$}\hldef{loglik,}
  \hlkwc{parameter}\hldef{=profile_table}\hlopt{$}\hldef{r,}
  \hlkwc{level}\hldef{=}\hlnum{0.95}\hldef{)}
\end{alltt}
\end{kframe}
\end{knitrout}
The construction of the confidence interval is best shown by a plot of the smoothed profile likelihood (Fig.~\ref{fig:mcap-plot}).
In this toy example, the exact likelihood can be calculated using the Kalman filter, and this is carried out by the \code{panelGompertzLikelihood} function.
The likelihood can then be maximized using a general-purpose optimization procedure such as \code{optim()} in R.
With large numbers of parameters, and no guarantee of convexity, this numerical optimization is not entirely routine.
One might consider a block optimization strategy, but here we carry out a simple global search, which took 15.5 minutes to compute the profile likelihood, once parallelized.
The deterministic search is also not entirely smooth, and so we apply MCAP as for the Monte Carlo search.
Both deterministic and Monte Carlo optimizations can benefit from a block optimization strategy which alternates between shared and unit-specific parameters \citep{breto19}.
Such algorithms can be built using the \panelPomp functions we have demonstrated, and they will be incorporated into the package once they have been more extensively researched.

\begin{knitrout}
\definecolor{shadecolor}{rgb}{0.969, 0.969, 0.969}\color{fgcolor}\begin{figure}

{\centering \includegraphics[width=5in]{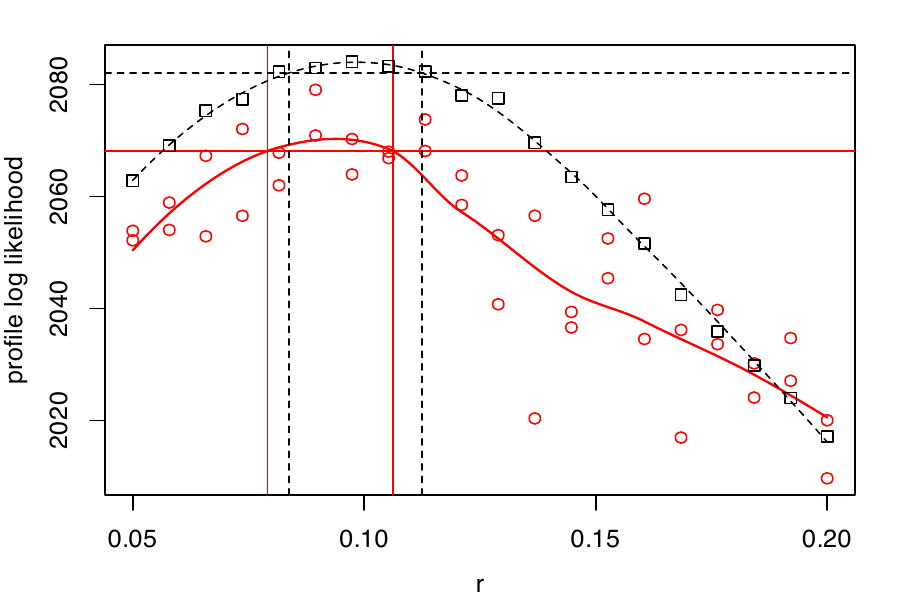} 

}

\caption[The Monte Carlo adjusted profile confidence interval (solid red lines, evaluation points shown as circles)]{The Monte Carlo adjusted profile confidence interval (solid red lines, evaluation points shown as circles). Construction using deterministic optimization of the likelihood calculated by the Kalman filter (dashed lines, evaluation points show as squares).}\label{fig:mcap-plot}
\end{figure}

\end{knitrout}

\section{Conclusion}

The analysis demonstrated in Section~\ref{sec:meth} gives one approach to plug-and-play inference for PanelPOMP models, but the scope of \panelPomp is far from limited to this approach.
\panelPomp is a general and extensible framework which encourages the development of additional functionality.
The \code{panelPomp} class and the corresponding workhorse functions provide an applications interface available to other future methodologies.
In this sense, \panelPomp provides an environment for sharing and developing PanelPOMP models and methods, both via future contributions to the \panelPomp package and via open source applications using \panelPomp.
This framework will facilitate comparison of new future methodology with existing methodology.

Likelihood evaluation and maximization was used to construct confidence intervals in Section~\ref{sec:meth}.
These calculations also provide a foundation for other techniques of likelihood-based inference, such as likelihood ratio hypothesis tests and model selection via Akaike's information criterion (AIC).
The examples discussed in Section~\ref{sec:examples} provide case studies in the use of these methods for scientific work.

Data analysis using large data sets or complex models may require considerable computing time.
Simulation-based methodology is necessarily computationally intensive, and access to a cluster computing environment extends the size of problems that can be tackled.
The workflow in Section~\ref{sec:meth} has a simple parallel structure that can readily take advantage of additional resources.
Embarrassingly parallel computations, such as computing the profile likelihood function at a grid of points, or replicated evaluations of the likelihood function, can be parallelized using the \pkg{foreach} package.

Panel data is widely available: for many experimental and observational systems it is more practical to collect short time series on many units than to obtain one long time series.
For time series data, fitting mechanistic models specified as partially observed Markov processes has found numerous applications for formulating and answering scientific hypotheses. \citep{breto09,king16}.
However, there are remarkably few examples in the literature fitting mechanistic nonlinear non-Gaussian partially observed stochastic dynamic models to panel data.
The \panelPomp package offers opportunities to remedy this situation.

The source code for \panelPomp is at \url{https://github.com/cbreto/panelPomp}.
Unit tests that cover 100\% of the code are provided at \url{https://github.com/cbreto/panelPomp/tests}, and these tests also provide useful examples of calls to the functions within \panelPomp.
The source code for this article is at \url{https://github.com/cbreto/panelPomp/vignettes/articles/package_tutorial}.

\section*{Acknowledgments}
The results in this paper were obtained using \R~4.4.1 with \pkg{panelPomp}~1.3.0 and pkg{pomp}~5.9.0.0.
This work was supported by National Science Foundation grants DMS-1761603 and DMS-1646108, National Institutes of Health grants 1-U54-GM111274 and 1-U01-GM110712, and by MCIN/AEI/10.13039/501100011033 grants PID2020-116242RB-I00 and PID2023-152348NB-I00

\bibliographystyle{jasa3}

\bibliography{bg}

\begin{thebibliography}{26}
\newcommand{\enquote}[1]{``#1''}
\expandafter\ifx\csname natexlab\endcsname\relax\def\natexlab#1{#1}\fi
\expandafter\ifx\csname url\endcsname\relax
  \def\url#1{{\tt #1}}\fi
\expandafter\ifx\csname urlprefix\endcsname\relax\def\urlprefix{URL }\fi

\bibitem[\protect\citeauthoryear{Asfaw, Park, King, and Ionides}{Asfaw et~al.}{2024}]{asfaw21}
Asfaw, K., Park, J., King, A.~A., and Ionides, E.~L. (2024), \enquote{A tutorial on spatiotemporal partially observed Markov process models via the R package spatPomp,} {\em arXiv:2101.01157\/}.

\bibitem[\protect\citeauthoryear{Bret\'{o}}{Bret\'{o}}{2014}]{breto14}
Bret\'{o}, C. (2014), \enquote{On idiosyncratic stochasticity of financial leverage effects,} {\em Statistics \& Probability Letters\/}, 91, 20--26.

\bibitem[\protect\citeauthoryear{Bret{\'o}}{Bret{\'o}}{2018}]{breto18}
Bret{\'o}, C. (2018), \enquote{Modeling and inference for infectious disease dynamics: a likelihood-based approach,} {\em Statistical Science\/}, 33, 57--69.

\bibitem[\protect\citeauthoryear{Bret\'{o}, He, Ionides, and King}{Bret\'{o} et~al.}{2009}]{breto09}
Bret\'{o}, C., He, D., Ionides, E.~L., and King, A.~A. (2009), \enquote{Time series analysis via mechanistic models,} {\em Annals of Applied Statistics\/}, 3, 319--348.

\bibitem[\protect\citeauthoryear{Bret{\'o}, Ionides, and King}{Bret{\'o} et~al.}{2020}]{breto19}
Bret{\'o}, C., Ionides, E.~L., and King, A.~A. (2020), \enquote{Panel data analysis via mechanistic models,} {\em Journal of the American Statistical Association\/}, 115, 1178--1188, \urlprefix\url{https://doi.org/10.1080/01621459.2019.1604367}.

\bibitem[\protect\citeauthoryear{Chambers}{Chambers}{1998}]{chambers98}
Chambers, J. (1998), {\em Programming with Data\/}, New York: Springer-Verlag.

\bibitem[\protect\citeauthoryear{Domeyer, Lee, Toyoda, Mehler, and Reimer}{Domeyer et~al.}{2022}]{domeyer22}
Domeyer, J.~E., Lee, J.~D., Toyoda, H., Mehler, B., and Reimer, B. (2022), \enquote{Driver-pedestrian perceptual models demonstrate coupling: implications for vehicle automation,} {\em IEEE Transactions on Human-Machine Systems\/}, 52, 557--566.

\bibitem[\protect\citeauthoryear{FitzJohn, Knock, Whittles, Perez-Guzman, Bhatia, Guntoro, Watson, Whittaker, Ferguson, Cori et~al.}{FitzJohn et~al.}{2020}]{fitzjohn20}
FitzJohn, R.~G., Knock, E.~S., Whittles, L.~K., Perez-Guzman, P.~N., Bhatia, S., Guntoro, F., Watson, O.~J., Whittaker, C., Ferguson, N.~M., Cori, A., et~al. (2020), \enquote{Reproducible parallel inference and simulation of stochastic state space models using odin, dust, and mcstate,} {\em Wellcome Open Research\/}, 5, 288.

\bibitem[\protect\citeauthoryear{Genolini}{Genolini}{2008}]{genolini08}
Genolini, C. (2008), \enquote{A (Not So) Short Introduction to \proglang{S}4,} \proglang{R} Foundation for Statistical Computing, \urlprefix\url{https://CRAN.R-project.org/doc/contrib/Genolini-S4tutorialV0-5en.pdf}.

\bibitem[\protect\citeauthoryear{He, Ionides, and King}{He et~al.}{2010}]{he10}
He, D., Ionides, E.~L., and King, A.~A. (2010), \enquote{Plug-and-play inference for disease dynamics: Measles in large and small towns as a case study,} {\em Journal of the Royal Society Interface\/}, 7, 271--283.

\bibitem[\protect\citeauthoryear{Ionides, Bret\'{o}, and King}{Ionides et~al.}{2006}]{ionides06}
Ionides, E.~L., Bret\'{o}, C., and King, A.~A. (2006), \enquote{Inference for nonlinear dynamical systems,} {\em Proceedings of the National Academy of Sciences of the USA\/}, 103, 18438--18443.

\bibitem[\protect\citeauthoryear{Ionides, Breto, Park, Smith, and King}{Ionides et~al.}{2017}]{ionides17}
Ionides, E.~L., Breto, C., Park, J., Smith, R.~A., and King, A.~A. (2017), \enquote{Monte {C}arlo profile confidence intervals for dynamic systems,} {\em Journal of the Royal Society Interface\/}, 14, 1--10.

\bibitem[\protect\citeauthoryear{Ionides, Nguyen, Atchad\'{e}, Stoev, and King}{Ionides et~al.}{2015}]{ionides15}
Ionides, E.~L., Nguyen, D., Atchad\'{e}, Y., Stoev, S., and King, A.~A. (2015), \enquote{Inference for dynamic and latent variable models via iterated, perturbed {B}ayes maps,} {\em Proceedings of the National Academy of Sciences of the USA\/}, 112, 719--724.

\bibitem[\protect\citeauthoryear{King, Nguyen, and Ionides}{King et~al.}{2016}]{king16}
King, A.~A., Nguyen, D., and Ionides, E.~L. (2016), \enquote{Statistical inference for partially observed {M}arkov processes via the {R} package pomp,} {\em Journal of Statistical Software\/}, 69, 1--43.

\bibitem[\protect\citeauthoryear{Lee, Chao, Lemaitre, Matrajt, Pasetto, Perez-Saez, Finger, Rinaldo, Sugimoto, Halloran, Longini, Ternier, Vissieres, Azman, Lessler, and Ivers}{Lee et~al.}{2020}]{lee20}
Lee, E.~C., Chao, D.~L., Lemaitre, J.~C., Matrajt, L., Pasetto, D., Perez-Saez, J., Finger, F., Rinaldo, A., Sugimoto, J.~D., Halloran, M.~E., Longini, I.~M., Ternier, R., Vissieres, K., Azman, A.~S., Lessler, J., and Ivers, L.~C. (2020), \enquote{Achieving coordinated national immunity and cholera elimination in {H}aiti through vaccination: A modelling study,} {\em The Lancet Global Health\/}, 8, e1081--e1089.

\bibitem[\protect\citeauthoryear{Marino, Peacor, Bunnell, Vanderploeg, Pothoven, Elgin, Bence, Jiao, and Ionides}{Marino et~al.}{2019}]{marino19}
Marino, J.~A., Peacor, S.~D., Bunnell, D.~B., Vanderploeg, H.~A., Pothoven, S.~A., Elgin, A.~K., Bence, J.~R., Jiao, J., and Ionides, E.~L. (2019), \enquote{Evaluating consumptive and nonconsumptive predator effects on prey density using field times series data,} {\em Ecology\/}, 100, e02583.

\bibitem[\protect\citeauthoryear{Marjoram, Molitor, Plagnol, and Tavar{\'e}}{Marjoram et~al.}{2003}]{marjoram03}
Marjoram, P., Molitor, J., Plagnol, V., and Tavar{\'e}, S. (2003), \enquote{Markov chain {M}onte {C}arlo without likelihoods,} {\em Proceedings of the National Academy of Sciences\/}, 100, 15324--15328.

\bibitem[\protect\citeauthoryear{Martinez-Bakker, King, and Rohani}{Martinez-Bakker et~al.}{2015}]{martinez-bakker15}
Martinez-Bakker, M., King, A.~A., and Rohani, P. (2015), \enquote{Unraveling the transmission ecology of polio,} {\em PLoS Biology\/}, 13, e1002172.

\bibitem[\protect\citeauthoryear{Michaud, de~Valpine, Turek, Paciorek, and Nguyen}{Michaud et~al.}{2021}]{michaud21}
Michaud, N., de~Valpine, P., Turek, D., Paciorek, C.~J., and Nguyen, D. (2021), \enquote{Sequential {M}onte {C}arlo methods in the nimble and nimbleSMC {R} packages,} {\em Journal of Statistical Software\/}, 100, 1–39.

\bibitem[\protect\citeauthoryear{Ning, Ionides, and Ritov}{Ning et~al.}{2021}]{ning21}
Ning, N., Ionides, E.~L., and Ritov, Y. (2021), \enquote{Scalable {M}onte {C}arlo inference and rescaled local asymptotic normality,} {\em Bernoulli\/}, 27, 2532--2555.

\bibitem[\protect\citeauthoryear{Ranjeva, Subramanian, Fang, Leung, Ip, Perera, Peiris, Cowling, and Cobey}{Ranjeva et~al.}{2019}]{ranjeva19}
Ranjeva, S., Subramanian, R., Fang, V.~J., Leung, G.~M., Ip, D.~K., Perera, R.~A., Peiris, J.~M., Cowling, B.~J., and Cobey, S. (2019), \enquote{Age-specific differences in the dynamics of protective immunity to influenza,} {\em Nature Communications\/}, 10, 1660.

\bibitem[\protect\citeauthoryear{Ranjeva, Baskerville, Dukic, Villa, Lazcano-Ponce, Giuliano, Dwyer, and Cobey}{Ranjeva et~al.}{2017}]{ranjeva17}
Ranjeva, S.~L., Baskerville, E.~B., Dukic, V., Villa, L.~L., Lazcano-Ponce, E., Giuliano, A.~R., Dwyer, G., and Cobey, S. (2017), \enquote{Recurring infection with ecologically distinct {HPV} types can explain high prevalence and diversity,} {\em Proceedings of the National Academy of Sciences of the USA\/}, 114, 13573--13578.

\bibitem[\protect\citeauthoryear{Romero-Severson, Volz, Koopman, Leitner, and Ionides}{Romero-Severson et~al.}{2015}]{romero-severson15}
Romero-Severson, E., Volz, E., Koopman, J., Leitner, T., and Ionides, E. (2015), \enquote{Dynamic variation in sexual contact rates in a cohort of {HIV}-negative gay men,} {\em American Journal of Epidemiology\/}, 182, 255--262.

\bibitem[\protect\citeauthoryear{Sisson, Fan, and Tanaka}{Sisson et~al.}{2007}]{sisson07}
Sisson, S.~A., Fan, Y., and Tanaka, M.~M. (2007), \enquote{Sequential {M}onte {C}arlo without likelihoods,} {\em Proceedings of the National Academy of Sciences\/}, 104, 1760--1765.

\bibitem[\protect\citeauthoryear{Wale, Jones, Sim, Read, and King}{Wale et~al.}{2019}]{wale19}
Wale, N., Jones, M.~J., Sim, D.~G., Read, A.~F., and King, A.~A. (2019), \enquote{The contribution of host cell-directed vs. parasite-directed immunity to the disease and dynamics of malaria infections,} {\em Proceedings of the National Academy of Sciences\/}, 116, 22386--22392.

\bibitem[\protect\citeauthoryear{Winsor}{Winsor}{1932}]{winsor32}
Winsor, C.~P. (1932), \enquote{The {G}ompertz curve as a growth curve,} {\em Proceedings of the National Academy of Sciences of the USA\/}, 18, 1--8.

\end{thebibliography}
\end{document}